# Manipulating 1-dimensinal skyrmion motion by external magnetic field gradient

Jaehun Cho[1, 2], Eiiti Tamura[2, 3, 4], Chaozhe Liu[5], Soma Miki[2, 3], Chun-Yeol You[6], June-Seo Kim[1], Hikaru Nomura[2, 3, 5], Minori Goto[2, 3], Ryoichi Nakatani[5], Yoshishige Suzuki[2, 3]

[1]Division of Nanotechnology, Daegu Gyeongbuk Institute of Science and Technology (DGIST), Daegu, Republic of Korea

[2]Graduate School of Engineering Science, Osaka University, Toyonaka, Osaka, Japan

[3]Center for Spintronics Research Network, Osaka University, Toyonaka, Osaka, Japan

[4]Department of Electronic Science and Engineering, Kyoto University, Kyoto, Kyoto, Japan

5Graduate School of Engineering, Osaka University, Suita, Osaka, Japan

[6]Department of Emerging Materials Science, Daegu Gyeongbuk Institute of Science and Technology (DGIST), Daegu, South Korea

**Abstract**

   We have investigated an analytic formula of the 1-dimensional magnetic skyrmion dynamics under external magnetic field gradient. We find excellent agreement between the analytical model and micromagnetic simulation results for various magnetic parameters such as the magnetic field gradient, Gilbert damping constant. We also observe much faster velocity of the chiral domain wall (DW) motion. The chiral DW is exist with smaller interfacial Dzyaloshinskii-Moriya interaction energy density cases. These results provide to develop efficient control of skyrmion for spintronic devices.

**Introduction**

In magnetic multilayer systems, the strong competition among Heisenberg exchange interaction, Dzyaloshinskii-Moriya (DM) interaction, and magnetocrystalline anisotropy can exhibit complex spin textures such as Skyrmions [1, 2], chiral magnetic domain walls (DWs) [3 - 5], Bloch lines [6, 7], and so on. A crucial contribution of interfacial DM interaction is directly related to a strong spin-orbit coupling at the interfaces between heavy metals and ferromagnets combined with the broken inversion symmetry at the interfaces [8 - 12]. These exotic spin textures based on DM interaction are topologically stable, which are appropriate for various applications as an information carrier and manipulator [13, 14]. Indeed, numerous theoretical and numerical studies have shown for the positive possibilities that magnetic skyrmions and chiral DWs could be essential ingredients for the next-generation spintronic devices for storage devices and logic application [8, 14]. A single or bunch of these topological objects are manipulated by laterally applied electrical currents due to spin transfer torque [15, 16] or spin-orbit torque [17, 18]. For the case below a critical current density, the magnetic textures are fastened due to a large pinning potential. When the electric current is larger than a critical current density, the non-negligible displacements of topological objects occur. While the electrical current injection technique is a promising method to drive multiple skrymions and chiral DWs synchronously, a large critical current density caused by poor resistivities of magnetic materials and the extremely narrow and long nanoscale wire architectures makes an insurmountable obstacle which is so-called "Joule heating problem" owing to Ohmic losses [19, 20]. Furthermore, the extra contributions such as Rashba effect and spin Hall effects lead to even more complex magnetization dynamics.

The magnetic field driven chiral DWs and magnetic solitons are received attentions because the system is totally governed by the the Landau–Lifshitz (LL) equation, which is equivalent to the the Landau–Lifshitz–Gilbert (LLG) equation when the magnetic damping constant of the system is small enough. Moreover, various manipulation idea such as DC or AC magnetic field driven magnetic solitons, the transverse magnetic field pulse induced DW and skyrmion racetrack are demonstrated recently [21, 22]. For realistic applications for information storage devices or logic applications, the magnetic skyrmion or DW racetrack should be compatible for the complementary metal-oxide-semiconductor (CMOS) architectures and the continuous miniaturization of CMOS architectures is essential for increasing the data capacity of the devices. For the magnetic field driven skyrmion or DW motions in real spintronic devices, the external magnetic field is applied from the ultrashort electrical current pulses passing through the conduction lines adjacent skyrmion or DW racetracks to minimize the energy consumptions. Naturally, the applied magnetic field to the magnetic racetracks are not uniform due to Oersted law.

In this work, the magnetic skyrmion and DW dynamics by applying gradient magnetic fields are systematically investigated by performing LLG simulations and Thiele approach. We described analytical and micromagnetic simulation studies of magnetic skyrmion dynamics in a 1-dimensional nanowire, force by magnetic field gradient along the z-direction while field gradient applied x-direction. According to the analytic model, the skyrmion dynamics in the nanowire with magnetic field gradient is proportionality to the skyrmion width and radius, and its dynamics in good agreement with the analytic model and micromagnetic simulation results. In micromagnetic simulations of DW dynamics, we observed much higher DW velocities than skyrmion one.

**Analytical model for magnetic field gradient driven skyrmions**

We briefly describe a simple theory for the skyrmion motion in our system. The motion of skyrmion in a two-dimensional film can be expressed by a Thiele's equations [23] for sufficiently slow varying and not too strong forces is fellow:

$$\mathbf{G} \times \dot{\mathbf{R}} + \alpha \mathcal{D} \dot{\mathbf{R}} = \mathbf{F} \quad (1)$$

Here, **R** is the center coordinate, **G** is the gyromagnetic coupling vector with the winding number of the skyrmion $q$, $\alpha$ is the Gilbert damping constant, $\mathcal{D}$ the dissipation dyadic and **F** the external force e.g. by electric currents, magnetic field gradients, and thermal fluctuations.

The gyromagnetic coupling vector **G** is given by,

$$\mathbf{G} = \int_V \hat{g}_{ij} dV. \text{ Here, } \hat{g}_{ij} = \frac{M_s}{|\gamma|} \mathbf{m} \cdot \left( \frac{\partial \mathbf{m}}{\partial x_i} \times \frac{\partial \mathbf{m}}{\partial x_j} \right),$$

$M_s$ is the saturation magnetization and $\gamma$ is the gyromagnetic ratio. The components of the dissipative force, which is second term of Eq. (1), $\alpha \mathcal{D}$ describes the friction of the skyrmion, $\mathcal{D} = \int_V d_{ij} dV$. Here, $d_{ij} = \int_V \frac{M_s}{|\gamma|} \frac{\partial \mathbf{m}}{\partial x_i} \cdot \frac{\partial \mathbf{m}}{\partial x_j} dV$.

We study the effects of a non-uniform perpendicular magnetic field with a longitudinal direction of nanowire. We neglect thermal fluctuation. As the skyrmion has a large magnetic moment relative to the ferromagnetic nanowire, the field gradient leads to a force acting on the skyrmion. The force is given by

$$\mathbf{F}_g = -\int H(r) \cdot \frac{\partial M}{\partial x_i} d^2 r$$

$$= -M_s \int H(r) \cdot \frac{\partial \mathbf{n}}{\partial x_i} d^2 r. \quad (2)$$

Because we apply the gradient magnetic field along the z-direction while field gradient applied x-direction, force by magnetic field gradient can be expressed as

$$\left( \mathbf{F}_g(h_g^z) \right)_x = -M_s h_g^z \int x \cdot \frac{\partial n_z}{\partial x} d^2 r$$

$$= M_s h_g^z \int_0^\infty dr\, r^2 \sin\Theta \frac{\partial \Theta}{\partial r} \int_0^{2\pi} d\varphi \cos^2\varphi$$

$$= \pi M_s h_g^z \int_0^\infty dr\, r^2 \sin\Theta \frac{\partial \Theta}{\partial r}$$

$$= 2\pi M_s h_g^z w^2 \int_0^\infty dt\, \frac{2t^2 \sinh^2(x)\sinh(t)\cosh(t)}{[\sinh^2(x)+\sinh^2(t)]^2}. \quad (3)$$

here, $h_g^z$ is perpendicular magnetic field gradient, $x = \frac{R}{w}$, $t = \frac{r}{w}$, where, $R$ is the radius of skyrmion, $w$ is width of the skyrmion, and $r$ is the position of magnetization.

Then, we consider the 1-dimensional equation of skyrmion motion in Eq. (1). Using magnetic field gradient force and dissipation above, we can calculate how the velocity of skyrmion depends on the magnetic field gradient as below

$$(\dot{R})_x = \frac{(F_g)_x}{\alpha D} = \frac{M_s h_g^z \int x \frac{\partial n_z}{\partial x} dx dy}{\frac{\alpha M_s}{|\gamma|} \int \frac{\partial \vec{n}}{\partial x} \cdot \frac{\partial \vec{n}}{\partial x} dx dy} \cong \frac{|\gamma|}{\alpha} \frac{h_g^z w^2 x^2}{x+\frac{q^2}{x}}. \quad (4)$$

For $q = \pm 1$ and $x \gg 1$, the velocity of skyrmion is simply calculated by

$$v_x \approx \frac{|\gamma|}{\alpha} h_g^z w R. \quad (5)$$

The value of $v_x$ increase with increasing skyrmion radius. The magnetic field gradient force has a large effect to large skyrmion because of large magnetic field differences from left and right side of skyrmion. Surprisingly, the skyrmion velocity is proportional to not only radius of skyrmion but also width of skyrmion. The skyrmion width and radius are determined as skyrmion shape which is correlated with magnetic parameters, thereby, the skyrmion width is not an independent variable in Eq. (5). Therefore, we find the relationship between the skyrmion velocities divided by skyrmion width ($v_x/w$) and the skyrmion radius as follows:

$$\left(\frac{v_x}{w}\right) \approx \frac{|\gamma|}{\alpha} h_g^z R \quad (6)$$

**Micromagnetic simulations**

The micromagnetic simulations were performed by using the MuMax3 which is numerically solve d the Landau-Lifshitz-Gilbert equation [24]. Here, we consider a nanowire shaped with 1000-nm-length, 100-nm-width, and 1.2-nm-thick. The discretized cell for simulations is set to be $2 \times 2 \times 1.2$ nm$^3$. In the simulation, a magnetic skyrmion is nucleated at the center of nanowire as shown in Fig. 1(a). As an example, here the magnetic parameters are saturation magnetization, $M_s$ = 560 kA/m, exchange stiffness constant, $A_{ex}$ = 12 pJ/m, perpendicular magnetic anisotropy energy $K$ = 1.1 MJ/m$^3$, interfacial Dzyaloshinskii-Moriya interaction energy density $D = 4.0$ mJ/m$^2$. To characterize the size and the shape of skyrmion, we take $m_z$ profiles across the center of skyrmion. We approximate the line profile across a skyrmion along the longitudinal direction of nanowire using a standard 360 ° domain wall profile [25, 26] as

$$m_z = \cos\left(2\arctan\left(\frac{\sinh\left(\frac{r-c}{w}\right)}{\sinh\left(\frac{R}{w}\right)}\right)\right). \quad (7)$$

Where $r$ is the position of magnetization, $c$ is the skyrmion center position, $w$ is width of the skyrmion and $R$ is the radius of skyrmion. The open black circles in Fig. 1(b) calculated values with the fitting parameter using Eq. 7. The obtained $R$ and $w$ are 17.56 nm and 3.54 nm, respectively. The material parameters in our simulations are chosen as table 1. The magnetic parameters selected for stable skyrmion conditions. In order to investigate skyrmion dynamics under applied magnetic field gradient, $h_g^z = (H_{\text{final}} - H_{\text{initial}})/L$, is applied the whole nanowire along $x$ - direction. We calculated the difference between the skyrmion center position at the initial time and the position at the final time as displacement. The final time is determined as the moment when the skyrmion stopped by nanowire edge. The determined skyrmion velocity is defined as the skyrmion displacement with respect to time.

**Magnetic field gradient driven skyrmions in nanowire**

The skyrmion width and radius are determined by magnetic parameters such as $M_s$, $A_{ex}$, $K$ and $D$, the skyrmion width and radius cannot consider separate. Since Eq. (5) implies the skyrmion velocity proportional to the skyrmion radius, however the skyrmion velocity are not matched well with skyrmion radius as shown in inset of Fig. 2. Because the values of skyrmion width are different with each skyrmions, the obtained skyrmion width are from 3.57 to 5.78 nm in the micromagnetic simulation results. The represented skyrmion radius dependence of the skyrmion velocities divided by skyrmion width ($v_x/w$) is plotted in Fig. 2. The black open circles indicate the simulation result for each skyrmion shown in table 1. The saturation magnetization of a whole skyrmion is 560 kA/m. The red line is the theoretical calculation with $\gamma = 176$ GHz/T, $\alpha = 0.3$, and $h_g^z = 10.0$ mT/μm. The value of $v_x/w$ increase with increasing skyrmion radius. The micromagnetic results are well matched as theoretical expectation results as shown in Eq. (6).

Based on the theoretical calculation, other magnetic parameters such as the damping constant and field gradient can also affect the skyrmion motion. In order to reveal the effects of skyrmion dynamics with various damping constant ($\alpha$) and field gradient ($h_g^z$), we perform micromagnetic simulations. Figure 3(a) and (b) indicate the $v_x/w$ with various $\alpha$ and $h_g^z$, respectively. In Fig. 3 (a) shows simulation results about the damping dependences of the skyrmion $v_x/w$ for the skyrmion radius (open symbols) along with the $v_x/w$ calculated with Eq. (6) (solid lines). The damping constant varied from 0.3 to 0.05 with $\gamma = 176$ GHz/T, and $h_g^z = 10.0$ mT/μm. We find that the $v_x/w$ value increase by decreasing the damping constant. The $v_x/w$ for various $h_g^z$ obtained by micromagnetic simulations (open symbols) and calculated by Eq. (6) (solid lines) as a function of skyrmion radius are displayed in Fig. 3 (b). The field gradient varied from 10.0 to 2.5 mT/μm with $\gamma = 176$ GHz/T, $\alpha = 0.05$. As the field gradient decrease, the value of $v_x/w$ decreases. The agreements between the results of micromagnetic simulation and Eq. (6) are excellent.

In Fig. 4, we have plotted the $v_x/w$ as a function of $\alpha$. Magnetic field gradient driven skyrmions dependence on the damping parameter is investigated at $h_g^z = 10.0$ mT/μm and skyrmion radius, $R = 23.30$ nm. Here the magnetic parameters are $M_s = 560$ kA/m, $A_{ex} = 12$ pJ/m, $K = 1.1$ MJ/m$^3$, and $D = 4.0$ mJ/m$^2$. The black open circles are micromagnetic simulation results and red solid line is calculated values using Eq. (6). The results show the inverse proportionality between $v_x/w$ and $\alpha$ which is consistent with Eq. (6). The decrease in alpha from 0.5 to 0.0125 resulted in an increase in $v_x/w$ from 0.07 to 1.79 GHz.

**Skyrmion dynamics for small interfacial DMI**

We set the DW center magnetization initially along the $y$ direction and relax it with magnetic field gradient, $h_g^z$ as 10.0 mT/μm. The iDMI energy density values coincides with the experimentally determined values [refs.]. The snapshot of the magnetizations with $M_s = 560$ kA/m, $A_{ex} = 12$ pJ/m, $K = 1.1$ MJ/m$^3$, $\alpha = 0.1$, and $D = 0.1$ mJ/m$^2$, depicted in Fig. 5(a) and the chiral DW displacements are shown in Fig. 5(b). As shown in colored lines in Fig. 5(b), the DW displacements are saturated after $t = 35$ ns for iDMI energy density is 0.1 mJ/m$^2$ and $t = 33$ ns for larger than 0.5 mJ/m$^2$.

**Summary**

In summary, we have investigated the skyrmion dynamics forced by a magnetic field gradient. The velocity of skyrmion is predicted analytically through the Thiele approach, which agrees well with micromagnetic simulation results. The skyrmion dynamics is related with skyrmion shape, Gilbert damping, magnetic field gradient. Interestingly, skyrmion velocities divided by skyrmion width is proportional to the skyrmion radius, magnetic field gradient and inverse Gilbert damping constant. For the DW dynamics case which is small iDMI energy density, DW velocity is much faster than the skyrmion velocities.

**Figure captions**

Fig. 1. (a) Spin structure of a hedgehog (Néel-type) skyrmion texture in a nanowire film. (b) Line profile across a skyrmion along the longitudinal direction of nanowire.

Fig. 2. The skyrmion radius dependence of skyrmion velocity divided by skyrmion width ($v_s/w$) with $\gamma$ = 176 GHz/T, $\alpha$ = 0.3, and magnetic field gradient $h_g^z$ = 1.00 mT/μm. inset: The skyrmion radius dependence of skyrmion velocity

Fig. 3. The skyrmion radius dependent of skyrmion motion velocity divided by skyrmion width ($v_s/w$) on the (a) damping constant, $\alpha$ and (b) magnetic field gradient $h_g^z$ dependence.

Fig. 4. The damping constant, $\alpha$ dependent of skyrmion motion velocity divided by skyrmion width ($v_s/w$) with $\gamma$ = 176 GHz/T, $R$ = 23.39 nm, and magnetic field gradient $h_g^z$ = 10.0 mT/μm.

Fig. 5. (a) The snapshots of the magnetization with various time. (b) DW displacements as a function of time with various iDMI energy densities. All lines are the DW displacements analyzed by $M_z/M_s$ values.

**Table caption**

Table 1. magnetic parameters for chosen stable skyrmion. The saturation magnetization of a whole skyrmions are 560 kA/m.

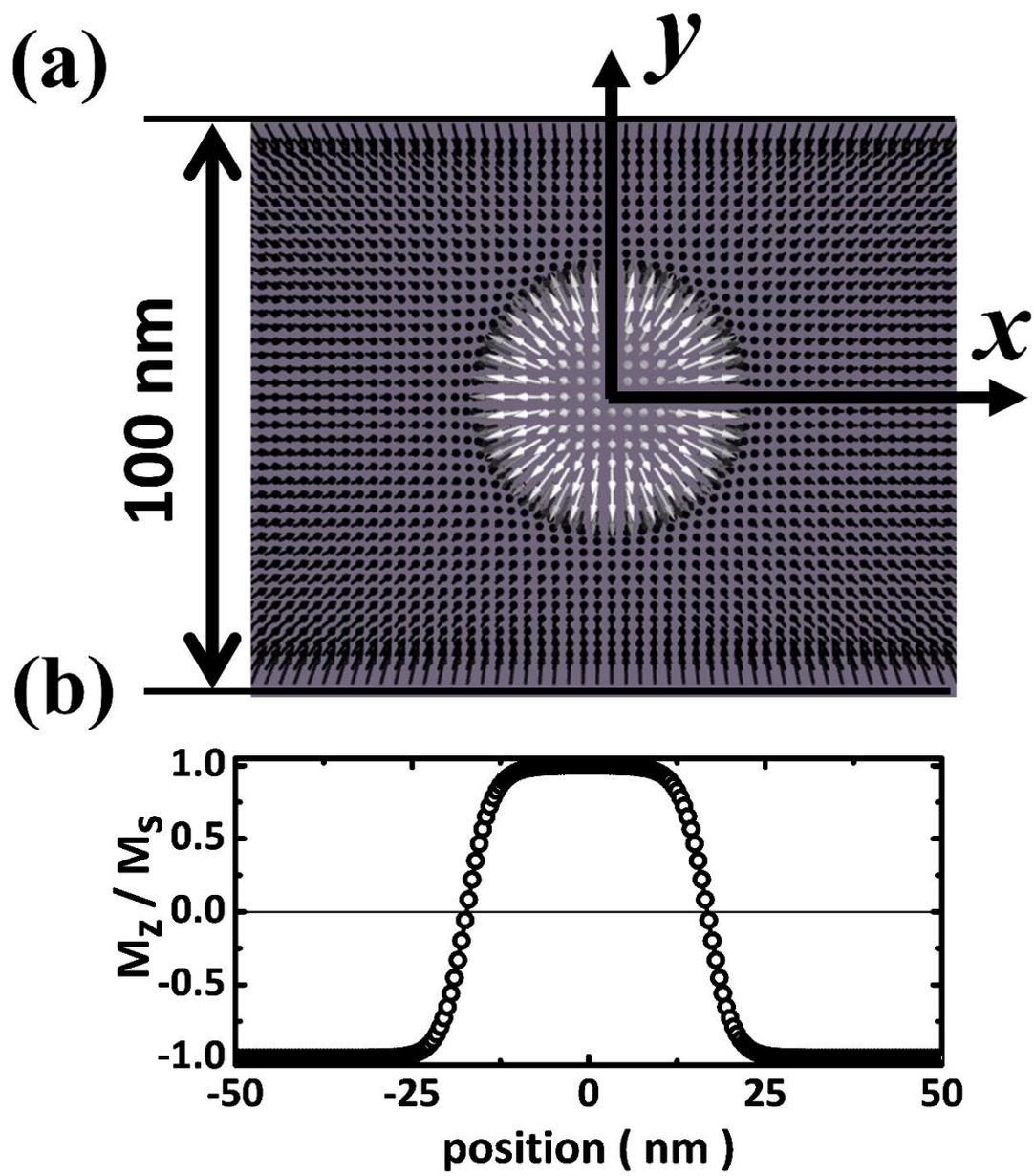

Figure 1.

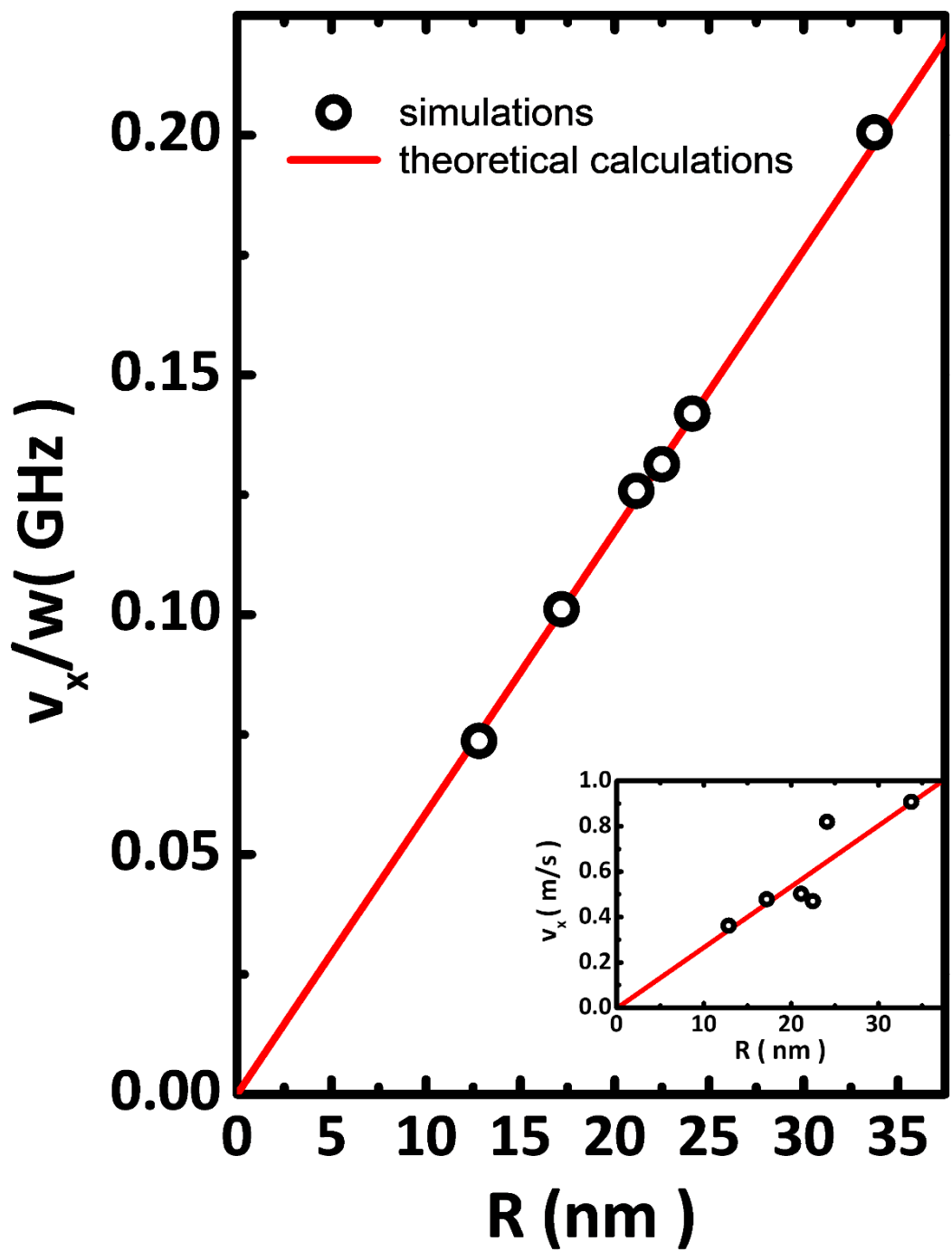

Figure 2.

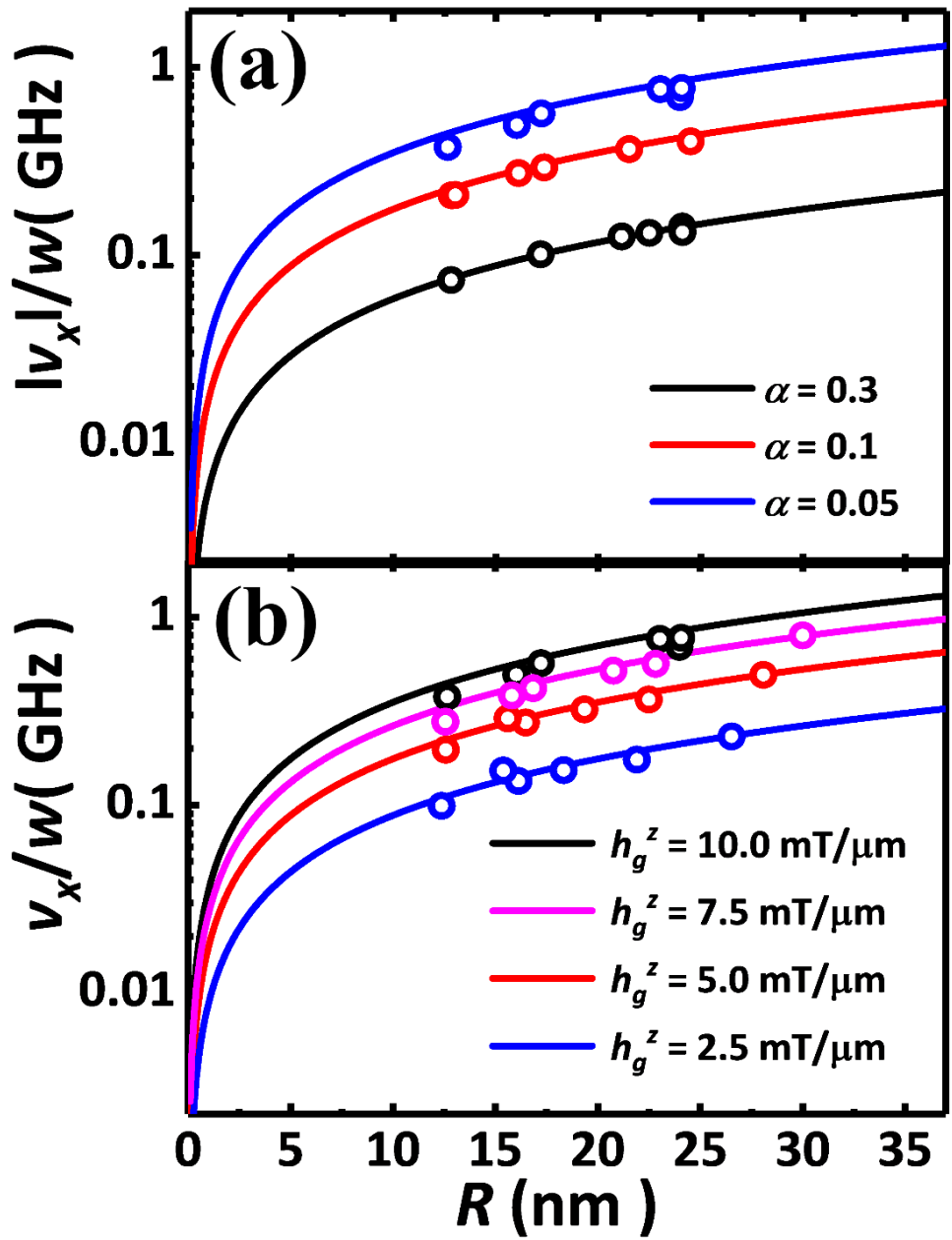

Figure 3.

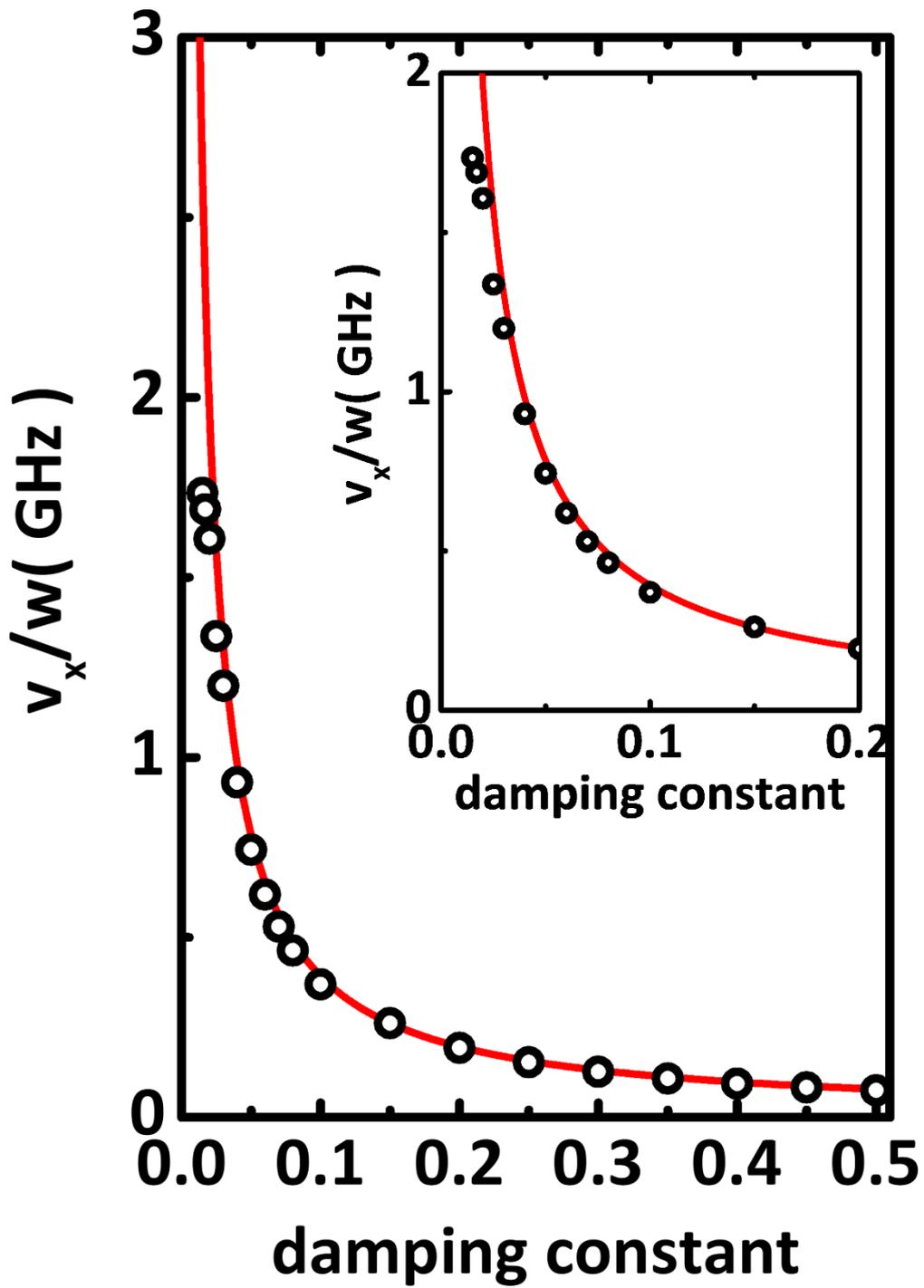

Figure 4.

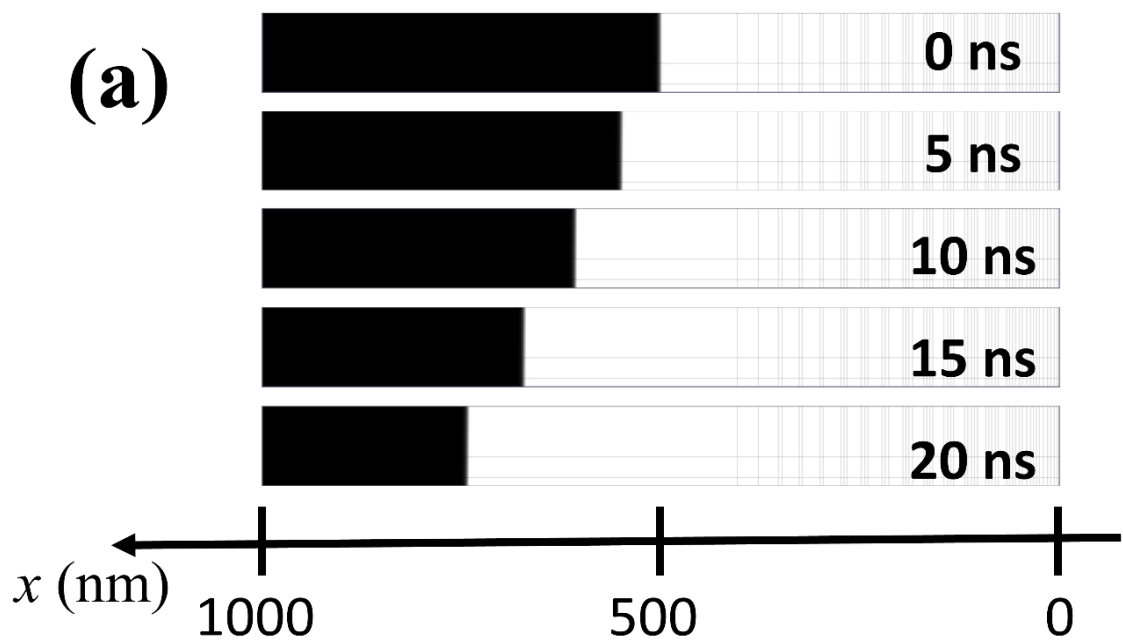
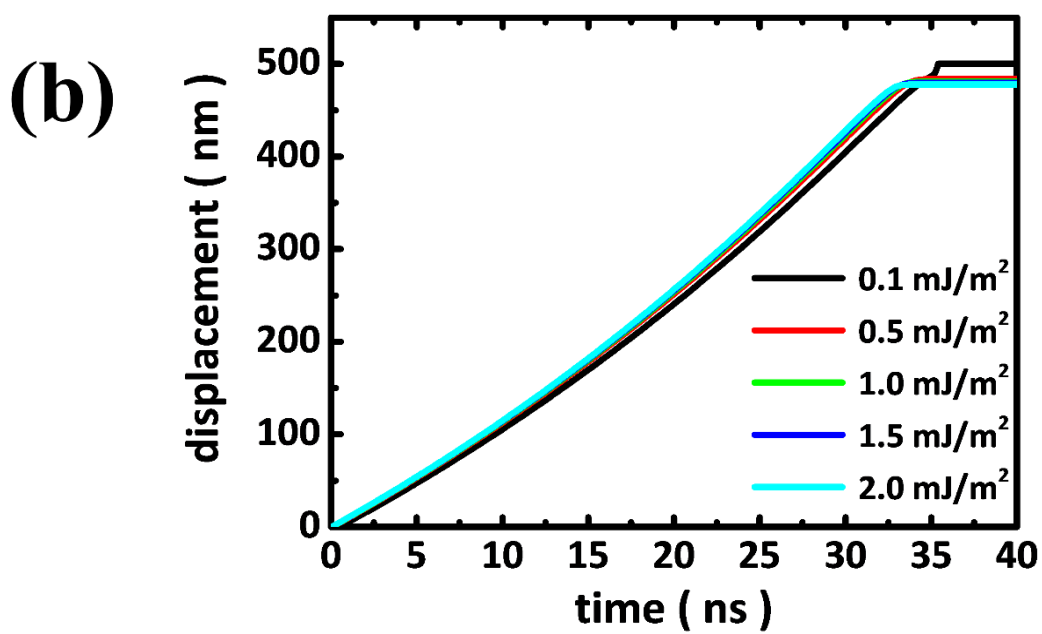

Figure 5.

| $A_{ex}$ (pJ/m) | $K$ (MJ/m$^3$) | $D$ (mJ/m$^2$) |
|---|---|---|
| 12 | 1.1 | 4 |
| 15 | 0.7 | 3 |
| 15 | 0.8 | 3.5 |
| 15 | 0.9 | 4 |
| 20 | 1.1 | 5 |
| 15 | 0.4 | 2 |

Table 1.